\documentclass[twocolumn,showpacs,preprintnumbers,amsmath,amssymb]{revtex4}

\usepackage[version=3]{mhchem} 
\usepackage{textcomp}

\usepackage{graphicx}
\usepackage{dcolumn}
\usepackage{bm}

\begin{document}

\author{V\'eronica Gavrilov-Isaac}
\affiliation{Sorbonne Universit\'es, UPMC Univ Paris 06, UMR 8234, PHENIX, F-75005 Paris, France
CNRS, UMR 8234, PHENIX, F-75005 Paris, France}
\email{veronica.gavrilov-isaac@upmc.fr}
\author{Sophie Neveu}
\affiliation{Sorbonne Universit\'es, UPMC Univ Paris 06, UMR 8234, PHENIX, F-75005 Paris, France
CNRS, UMR 8234, PHENIX, F-75005 Paris, France}
\author{Vincent Dupuis}
\affiliation{Sorbonne Universit\'es, UPMC Univ Paris 06, UMR 8234, PHENIX, F-75005 Paris, France
CNRS, UMR 8234, PHENIX, F-75005 Paris, France}
\author{Delphine Talbot}
\affiliation{Sorbonne Universit\'es, UPMC Univ Paris 06, UMR 8234, PHENIX, F-75005 Paris, France
CNRS, UMR 8234, PHENIX, F-75005 Paris, France}
\author{Val\'erie Cabuil}
\affiliation{Sorbonne Universit\'es, UPMC Univ Paris 06, UMR 8234, PHENIX, F-75005 Paris, France
CNRS, UMR 8234, PHENIX, F-75005 Paris, France}

\title{Synthesis of Fe$_3$O$_4$@CoFe$_2$O$_4$@MnFe$_2$O$_4$ trimagnetic core/shell/shell nanoparticles}

\maketitle


Magnetic nanoparticles with spinel structure MFe$_2$O$_4$ (M = Fe, Co, Mn, Zn, Ni, Cu...) have been extensively studied for their various magnetic applications ranging from magnetic energy storage to biomedical applications.\cite{Frey2009}$^,$\cite{Lu2007} Different synthesis methods, such as co-precipitation\cite{Tourinho1990}$^,$\cite{Neveu2002} $,$ forced hydrolysis in a polyol medium\cite{Ammar2001}$^,$\cite{Caruntu2002}$,$ micro-emulsions\cite{Moumen1996}$,$ hydrothermal synthesis\cite{Daou2006}$^,$\cite{Horner2009}$,$ microfluidic process\cite{AbouHassan2012}$,$ or thermal decomposition\cite{Perez-Mirabet2013}$^,$\cite{Song2004}$,$ have been used to control size, shape and composition of these nanomaterials. Thermal decomposition of metal precursors has been demonstrated to be a very effective method to prepare monodisperse nanoparticles with controlled morphology\cite{Sun2004} . To develop original magnetic properties bimagnetic core/shell nanostructured particles have been synthesized and characterized.\cite{Masala2006} These particles are a combination of a magnetic hard phase (e.g. CoFe$_2$O$_4$) and a magnetic soft phase (e. g. MnFe$_2$O$_4$, ZnFe$_2$O$_4$ or Fe$_3$O$_4$), and possess unique magnetic properties.\cite{Song2012}. They are expected to have a good efficiency for magnetic hyperthermia.\cite{Lee2011}

We report here the  synthesis and characterization of what we call trimagnetic core/shell/shell Fe$_3$O$_4$@CoFe$_2$O$_4$@MnFe$_2$O$_4$ nanoparticles. These particles are a combination of a hard phase (CoFe$_2$O$_4$) and two soft phases (Fe$_3$O$_4$ and MnFe$_2$O$_4$), and have unique magnetic characteristics. The Fe$_3$O$_4$ core particles were synthesized according to the procedure described by Sun and all\cite{Sun2004} by high-temperature decomposition ($\sim$$280$\textcelsius) of a mixture of Fe(acac)$_3$, oleic acid, oleylamine, 1,2-hexadecanediol and benzyl ether. To synthesize Fe$_3$O$_4$@CoFe$_2$O$_4$ core/shell and Fe$_3$O$_4$@CoFe$_2$O$_4$@MnFe$_2$O$_4$ core/shell/shell nanoparticles, a seed-mediated growth at high temperature method was used.
The Fe$_3$O$_4$ nanoparticles seeds  (1.5 mmol) dispersed in heptane were mixed under a flow of nitrogen with a mixture of Fe(acac)$_3$ (1 mmol), Co(acac)$_2$ (0.5 mmol), oleic acid (6 mmol), oleylamine (6 mmol), 1,2-hexadecanediol (10 mmol), benzyl ether (20 mL). The solution was first heated to $100$\textcelsius \ for 30 min to remove heptane, then to reflux ($\sim$$300$\textcelsius) for 1h. The final mixture was cooled down to room temperature, washed with ethanol and a black precipitate was collected after magnetic precipitation. The separated nanoparticles were re-dispersed in heptane, and a black ferrofluid composed of Fe$_3$O$_4$@CoFe$_2$O$_4$ bimagnetic core@shell nanoparticles was produced.
Under the same conditions, Fe$_3$O$_4$@CoFe$_2$O$_4$@MnFe$_2$O$_4$ core/shell/shell nanoparticles dispersed in heptane, were obtained by mixing the Fe$_3$O$_4$@CoFe$_2$O$_4$ bi-magnetic seeds (1.5 mmol) with a mixture made of 1 mmol of Fe(acac)$_3$ and 0.5 mmol of Mn(acac)$_2$.

\begin{figure}[h]
\includegraphics[width=8.5cm]{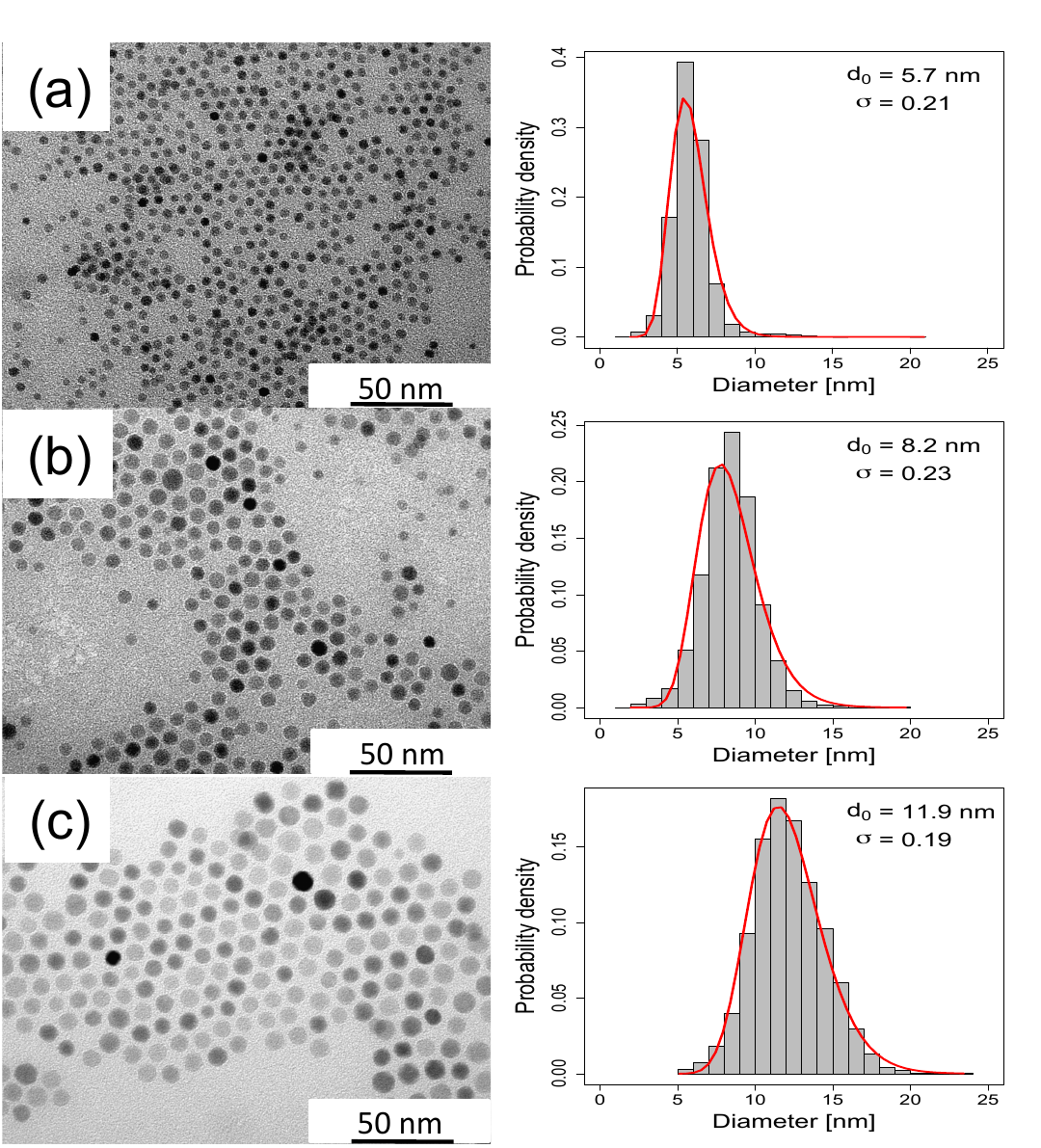}
\caption{\label{Figure1}TEM images and size distribution histograms of (a) 6 nm Fe$_3$O$_4$ core nanoparticles, (b) 8 nm Fe$_3$O$_4$@CoFe$_2$O$_4$ core/shell nanoparticles and (c) 12 nm Fe$_3$O$_4$@CoFe$_2$O$_4$@MnFe$_2$O$_4$ core/shell/shell nanoparticles obtained with a JEOL 100CX (x93000).}
\end{figure}

Figure 1 shows the transmission electron microscopy (TEM) images of 6 nm Fe$_3$O$_4$ core, 8 nm Fe$_3$O$_4$@CoFe$_2$O$_4$ core/shell, and 12 nm Fe$_3$O$_4$@CoFe$_2$O$_4$@MnFe$_2$O$_4$ core/shell/shell nanoparticles.
TEM size analysis indicates that particles are monodisperse with narrow size distributions. Histograms of core, core/shell, and core/shell/shell nanoparticles provide a nice illustration of the progressive increase of particles size as soon as a new magnetic shell is added.

\begin{figure}[h]
\centering
\includegraphics[scale=0.4]{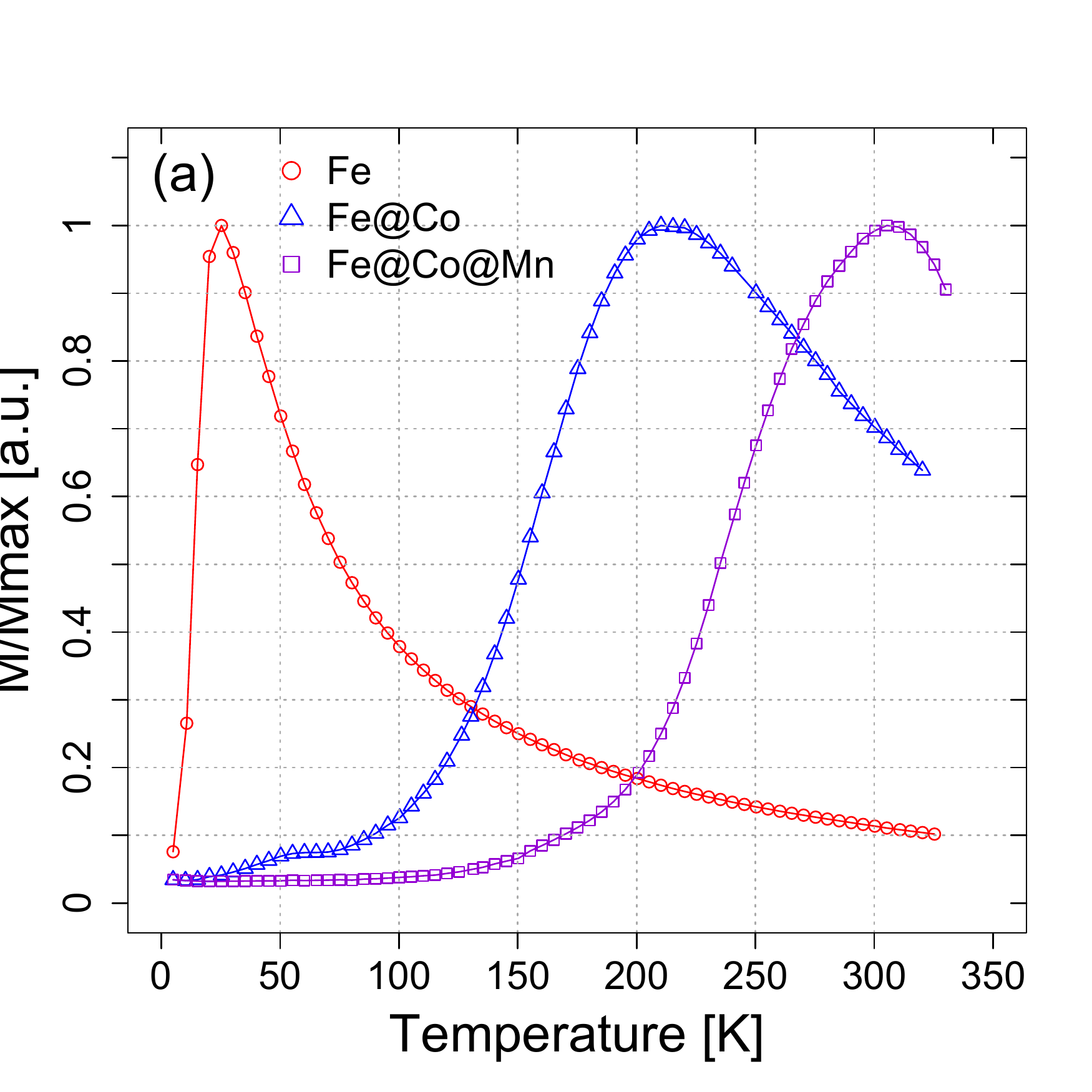}
\includegraphics[scale=0.4]{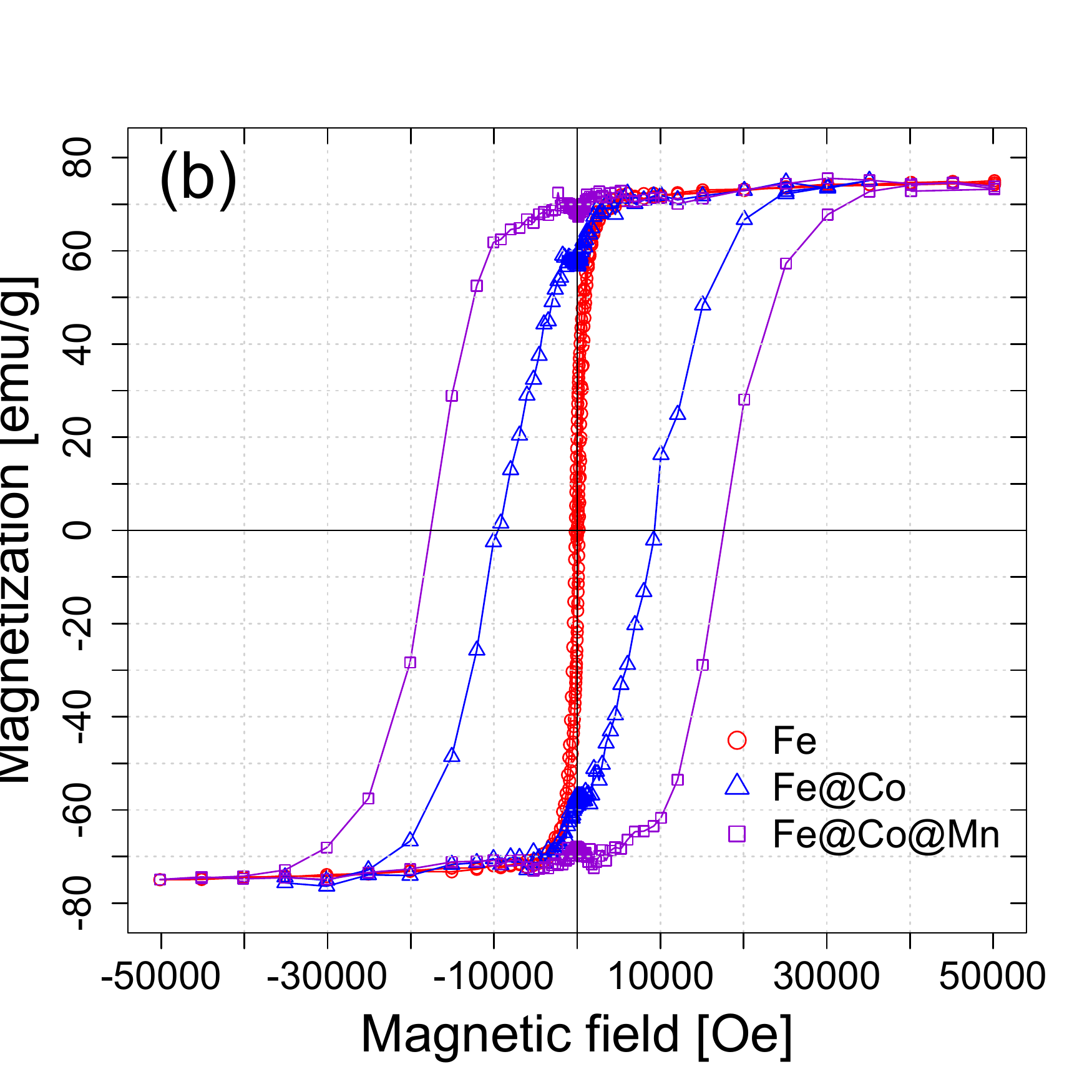}
\caption{(a) Blocking temperature and (b) coercivity at 5K of Fe$_3$O$_4$ core nanoparticles (Fe), Fe$_3$O$_4$@CoFe$_2$O$_4$ core/shell nanoparticles (Fe@Co) and Fe$_3$O$_4$@CoFe$_2$O$_4$@MnFe$_2$O$_4$ core/shell/shell nanoparticles (Fe@Co@Mn).}
\label{Figure3(A)}
\end{figure}

We compare the magnetic properties (blocking temperature and coercivity) of the core, core/shell and core/shell/shell particles. Figure 2a shows the zero-field cooled (ZFC) temperature dependence of magnetization under a 50 Oe field. The blocking temperature (T$_B$) increases when comparing core, core/shell, and core/shell/shell structures. Fe$_3$O$_4$ nanoparticles display a blocking temperature at 25 K, although this of  Fe$_3$O$_4$@CoFe$_2$O$_4$ is around 210 K. The increase between the blocking temperatures of the core/shell and core/shell/shell structures (T$_B$ = 305 K) is lower, indicating that the magnetic hard phase shell (CoFe$_2$O$_4$) has a more important impact on the blocking temperature compared to magnetic soft phase shell (MnFe$_2$O$_4$).

Magnetization as a function of the magnetic field acquired at 5K, is displayed in Figure 2b. The temperature is lower that the blocking temperature and a hysteresis look is obtained for each sample. This result is quite different of the two phase magnetic behavior that  would have been obtained with physically mixed CoFe$_2$O$_4$ and MnFe$_2$O$_4$ nanocrystals\cite{Song2012}. This confirmes the core/shell and core/shell/shell structures of the synthesized particles. 
Coercivity H$_C$ is significantly different in bimagnetic core/shell and trimagnetic core/shell/shell nanoparticles compared to magnetic core nanoparticles.
Hysteresis measurements show that coercivity increases when the magnetic soft phase Fe$_3$O$_4$ core is coated with a magnetic hard phase CoFe$_2$O$_4$ shell. It changes from 0.2 kOe for Fe$_3$O$_4$ nanoparticles to 9 kOe for Fe$_3$O$_4$@CoFe$_2$O$_4$ nanoparticles.
These results regarding the core/shell particles are in good accordance with those of Song and Zhang\cite{Song2012} who have evidenced a coercivity increase for MnFe$_2$O$_4$ particles coated with a CoFe$_2$O$_4$ shell and a decrease for CoFe$_2$O$_4$ particles coated by a MnFe$_2$O$_4$ shell. In their paper, the authors discussed their observations in terms of a simple model in which coercitivity is ruled by the proportion of hard and soft phases within a particle. Our results, for a Fe$_3$O$_4$@CoFe$_2$O$_4$ core/shell particles coated with an additional shell made of a magnetic soft phase (here MnFe$_2$O$_4$ but similar results were obtained for a second shell made of Fe$_3$O$_4$), show that contrary to expectations from this simple model, the coercivity is increased (H$_c$ = 17 kOe). This shows that the physics governing the magnetic properties of trimagnetic core/shell/shell nanoparticles  is certainly more complex than anticipated from the results on bimagnetic core/shell nanoparticles and should be investigated more thoroughly by numerical simulations and on the experimental side by varying the shell thicknesses and the nature of materials. In the same time, it provides new opportunities towards a fine tuning of the magnetic anisotropy of magnetic nanoparticles.

\begin{acknowledgements}

The author thanks Aude Michel for the technical assistance, and P. Beaunier for the access to the TEM platform.

\end{acknowledgements}

\bibliography{Manuscript_Gavrilov-Isaac}

\end{document}